\documentclass[fleqn,usenatbib]{mnras}

\usepackage{newtxtext,newtxmath}

\usepackage[T1]{fontenc}

\DeclareRobustCommand{\VAN}[3]{#2}
\let\VANthebibliography\thebibliography
\def\thebibliography{\DeclareRobustCommand{\VAN}[3]{##3}\VANthebibliography}

\usepackage{graphicx}	
\usepackage{amsmath}	
\usepackage{float}

\def\las{\mathrel{\hbox{\rlap{\hbox{\lower3pt\hbox{$\sim$}}}\hbox{\raise2pt\hbox{$<$}}}}}

\title[LT Survey]{A Dedicated Lunar Trojan Asteroid Survey with Small Ground-Based Telescopes}

\author[C. R. Gregg and P. A. Wiegert]{Cole R. Gregg,$^{1,2}$\thanks{E-mail: cgregg2@uwo.ca} Paul. A. Wiegert$^{1,2}$
\\
$^{1}$Dept. of Physics and Astronomy, The University of Western Ontario London, Canada\\
$^{2}$Institute for Earth and Space Exploration (IESX), The University of Western Ontario London, Canada
}

\date{Accepted 2022 January 17. Received 2022 January 17; in original form 2021 October 26}

\pubyear{2022}

\begin{document}
\label{firstpage}
\pagerange{\pageref{firstpage}--\pageref{lastpage}}
\maketitle

\begin{abstract}
A co-orbital asteroid shares the orbit of a secondary body about its primary. Though more commonly encountered as an asteroid that shares a planet's orbit around the Sun, a co-orbital asteroid could similarly share the orbit of the Moon around the Earth. Though such asteroids would be close to Earth and so relatively bright, their rapid on-sky motion is such that they might escape detection by near-Earth asteroid surveys. The discovery of such lunar co-orbital asteroids (which we will refer to generically here as Lunar Trojans or LTs) would advance our understanding of inner Solar System orbital dynamics and would provide research opportunities for the growing number of missions slated for cislunar space. No LT asteroids are currently known and the last published survey dedicated to these asteroids was conducted nearly 40 years ago. It has been theoretically determined that orbits near the Earth-Moon L4 and L5 points could survive for several million years. Although this timescale is shorter than the lifetime of the Solar System, it introduces the possibility of the temporary capture of asteroids into the LT state. This project aims to observationally evaluate the population of LTs with modern techniques. Using four small ground-based telescopes from the iTelescope network, $8340\;deg^2$ on the sky were surveyed down to $15^{th}$ magnitude. Though one fast-moving near-Earth object was detected, no LTs were observed. We deduce an upper limit of $\lesssim 5$ LTs with $H<26$.

\end{abstract}

\begin{keywords}
minor planets, asteroids, general < Planetary Systems, surveys < 
Astronomical Data bases, planets and satellites: detection < Planetary 
Systems, Moon < Planetary Systems

\end{keywords}



\section{Introduction}










Ground-based, telescopic small body surveys have become increasingly sophisticated over the last few decades, and are the primary mechanism for the discovery and study of the near-Earth asteroid population. There are currently over 26,000 known near-Earth asteroids (NEAs) according to the International Astronomical Union's (IAU) Minor Planet Center (MPC)\footnote{Available online at: \url{https://minorplanetcenter.net/mpc/summary}}. Of these, 2,255 are potentially hazardous asteroids (PHAs, indicating that it poses a hypothetical threat of impact, having an Earth minimum orbital intersection distance (MOID) less than 0.05 [$AU$] and an absolute magnitude $H<22.0$; \cite{Atkinson_2000}; \cite{Stokes_2003}), 940 have diameters $>1\;[km]$, and as a population present a non-zero risk to our planet. 

The value of asteroid/comet surveys goes beyond planetary defense, serving to further our scientific understanding of these rocky/icy bodies as well. Asteroids and comets are the leftover remnants from the earliest eras of the Solar System and therefore can provide us with clues about its formation. Interactions with small bodies likely played an important role in the evolution of Earth and its biosphere. The valuable information these bodies hold about early times is not otherwise easily obtainable.

This project consists of a sky survey directed at asteroids that are co-orbital with the Moon. Co-orbital bodies can be of several types (discussed further below), of which the Trojan asteroid subclass is the best known. For simplicity we will refer to all lunar co-orbitals as Lunar Trojans (LTs) here, as searches for all lunar co-orbital types proceed along very much the same lines. No asteroid of this type has been discovered to date. If a LT were to be discovered it would be, and remain, roughly at the lunar distance (LD) from Earth, providing researchers with the opportunity to do a long-term study on a nearby asteroid, potentially answering questions such as whether asteroids played a role in bringing the necessities of life, like water or organic molecules, to our planet, or in the Late Heavy Bombardment \citep{cukgla09}.

The discovery of a LT would be relevant to the successors to NASA's Asteroid Redirect Mission (ARM), also known as the Asteroid Retrieval and Utilization (ARU) mission. The mission concept had the objective of robotically returning a multi-ton boulder from a large NEA to cislunar space, with the goals of enabling resource utilization; human exploration; enhancing NEA detection, tracking, and characterization for planetary defense; and demonstrating planetary defense techniques \citep{mazanek_2015}.LTs are much closer and have lower delta-v requirements than typical NEAs. 
The discovery of a LT would also permit asteroid resource utilization without the need for a retrieval mission, and enabling space mining opportunities such as those explored by the Trans Astronautica (TransAstra) Corporation \footnote{See \url{https://www.transastracorp.com/}} (e.g. mining for water-ice to enable self-sustaining space missions).
LTs would also provide scientific targets and resources to the planned lunar orbiting space station Lunar Gateway, now known as The Gateway, a part of the Artemis missions \citep{Gateway}, making surveys for LTs particularly timely.

\subsection{Trojan and Co-orbital Dynamics}

The special solutions of the three-body problem resulting in the 5 Lagrange points have been well studied since their discovery by Leonhard Euler and Joseph-Louis Lagrange in the $18^{th}$ century (\cite{euler_1767} and \cite{lagrange_1772}). The 3 collinear Lagrange points (L1, L2, and L3) are saddle points of the total potential (gravitational + centrifugal) in the rotating frame. Particles displaced slightly from these points will continue to move away; these 3 locations are unstable. The triangular Lagrange points lie $\pm60^\circ$ from the secondary body about its orbit from the perspective of the primary body (L4 and L5 respectively). A particle from these locations perturbed slightly can librate about these points and remain stable indefinitely \citep{murder99}.

The Lagrange points are just the beginning of the discussion of co-orbital objects that are in a 1:1 mean-motion resonance with larger bodies in the Solar System. Material that librates about these stable L4 and L5 positions are considered to be on {\it tadpole} orbits due to their shape in the rotating frame at the orbital frequency of the two massive bodies. Material can also be trapped in {\it horseshoe} orbits (also named from their shape in the rotating frame) encircling the L3, L4 and L5 positions \citep{murder99}. {\it Retrograde} or {\it quasi-satellites} have motions which resemble that of satellites, however unlike traditional satellites, quasi-satellites lie outside of the Hill Sphere of the primary body and are unstable in the inner Solar System \citep{Mikkola_Innanen_1997}. Additionally,  {\it compound} co-orbital states which include one or more of these types of motion and which may move between states over time are also possible \citep{Wiegert1997}. 

Each of these co-orbital states could exist in the Earth-Moon system as well. However, the on-sky motion of these different states would be very similar and the differences only apparent after long times (usually tens to hundreds of orbits). As a result we will not make fine distinctions between these states here and refer to them collectively as Lunar Trojans (LTs). 

\subsection{Trojan Populations in the Solar System}

The Jupiter Trojan asteroids in the Sun-Jupiter system are the most famous co-orbital population in our Solar System. This large (thousands) population of asteroids at the L4 and L5 positions can drift up to $40^\circ$ from the nominal L points \citep{Erdi_2013}. However, the Jupiter Trojans are not the only known population of Trojan asteroids. As of June 2021 the MPC lists 28 Neptune Trojans, 9 Martian Trojans, 1 Earth Trojan, and 1 Uranus Trojan (see \cite{Alexandersen_2021} for a review).

Though no bodies are known at the Earth-Moon triangular points, the presence of a meteoric population there was proposed by \cite{moulton_1900}, and their presence has been debated since \cite{kordylewski_1961} first claimed their observation. These so-called Kordylewski dust clouds are so faint that proving their existence has been rather difficult. \cite{Balogh_2018} concluded -on the basis of polarimetric observations- that they do exist despite the Japanese Hiten space probe not finding an increase in the concentration of dust compared to the surrounding space when passing through the Earth-Moon L4 and L5 points \citep{Igenbergs_1991}.

An important motivation for this study is that there is theoretical evidence that LTs could exist under the right conditions. \cite{LisCha08} used numerical integrations to determine that orbits near the Earth-Moon L4 and L5 points can survive for several million years, even when solar perturbations and the far smaller perturbations from other planets are included. On astronomical timescales these orbits are unstable, suggesting that if any object were to be present now, it most likely entered the region in the relatively recent past. \cite{LisCha08} findings thus introduces the possibility of the Earth-Moon system capturing a temporary lunar co-orbital asteroid, or a LT, for a long period of time (humanly speaking), that would likely be in a tadpole or horseshoe orbit with the Moon about the Earth.

\subsubsection{Minimoons}

There is another class of minor bodies, originally predicted by \cite{Chant_1913} and \cite{Denning_1917}, that should be included in the discussion of near-Earth space: temporarily captured orbiters (TCOs) or minimoons. As asteroids pass near Earth their orbits can be altered by the Earth or the Moon's gravitational attraction, a fraction can be captured as minimoons for a few orbits \citep{Clark_2016}. There have been 2 observed minimoons of Earth, 2006 RH$_{120}$ \citep{Kwiatkowski_2009} and 2020 CD$_3$ \citep{Fedorets_2020}. Although minimoons are not confined to the lunar orbit, it is possible for one to be observed when surveying for LTs.

\subsection{Survey Motivation}

The most recent published survey dedicated to the search for LTs was conducted by \cite{ValFre83} nearly 40 years ago. They executed a photographic search of the 5 Earth-Moon Lagrangian positions (and surrounding areas), as well as the L2 position in the Sun-Earth system. Their observations were obtained through a 0.61-meter telescope at the Warner and Swasey Observatory. Their data spanned $60^\circ$ along the lunar orbital path by $5^\circ$ around the L5 position, and $48^\circ$ x $5^\circ$ around L4. Three pairs of photographic images were taken for the L4 and L5 positions tracking at different rates: the first of which was sidereal, then the telescope tracked the predicted stable libration orbital positions as computed by the team, and then at the lunar rate. Each plate was scanned using a stereoscopic microscope and light table. For a few plates, they were able to digitize the images and run an automated search. Despite their efforts, no natural or artificial objects were found. 

Since the \cite{ValFre83} search, technology has pushed forward our ability to conduct such surveys. A few of the major surveys currently operating include: Spacewatch, the Catalina Sky Survey, the Panoramic Survey Telescope And Rapid Response System (Pan-STARRS), and the Asteroid Terrestrial-impact Last Alert System (ATLAS). Importantly, not all of these surveys routinely search for objects moving as fast as LTs, which move on average at $13^\circ/day$ (discussed later in this section). As a result, LTs may have gone unobserved despite the unprecedented telescopic coverage of near-Earth space modern surveys provide.


The Spacewatch project employs both a 0.9-meter and a 1.8-meter telescope. Searches for near-Earth objects (NEOs, indicating that the objects perihelion distance lies within 1.3 [$AU$] of the Sun) are performed for on-sky motions between 0.05 and $2.5^\circ/day$, covering 1400 $deg^2$ of sky each lunation on average, usually concentrated on the ecliptic \citep{Spacewatch_2006}. The Catalina Sky Survey uses their 0.68-meter and 1.5-meter telescopes in their search for PHAs, with a 1.0-meter follow-up telescope. This survey has been one of the most successful in its operation, having discovered nearly half of the known NEO population \citep{Christensen2018StatusOT}. We have not been able to find the range of on-sky angular velocities the Catalina Sky Survey is able to detect in the literature, however, based upon NEA discoveries such as 2014 AA, they are able to reach at least $4.5^\circ/day$ \citep{Catalina16}. Pan-STARRS is able to search the entire sky north of $-30^\circ$ in declination multiple times a year with the use of two 1.8-meter telescopes. They are able to detect objects with on-sky angular velocities from 0.3 to $0.7^\circ/day$ and 1.2 to $5^\circ/day$ \citep{PanSTARRS13}. ATLAS also has a primary goal of finding PHAs using their 0.5-meter telescopes. Due to their large field of view (FoV) and short $30s$ exposures, ATLAS is able to detect almost any asteroid brighter than their sensitivity limit at a declination north of $-35^\circ$, including objects moving as fast as $50^\circ/day$ at their closest approach \citep{ATLAS18}.

Although current sky surveys searching for NEOs are very capable surveys, the inherent high-speed motion of LTs may allow them to elude these searches. The Moon traverses the night sky at an average angular velocity of $13.2^\circ/day$. If an asteroid is co-orbital with the Moon about the Earth, its expected average on-sky angular velocity would also be $13.2^\circ/day$. If however, the asteroid is in a tadpole orbit librating about these points, or a horseshoe orbit, enclosing these points, there would be a range of motion these bodies could have centred on this average, which will be examined next.

\subsubsection{Simulated On-Sky Motion} \label{sec:Simulation}

\begin{figure}
	\includegraphics[width=\columnwidth]{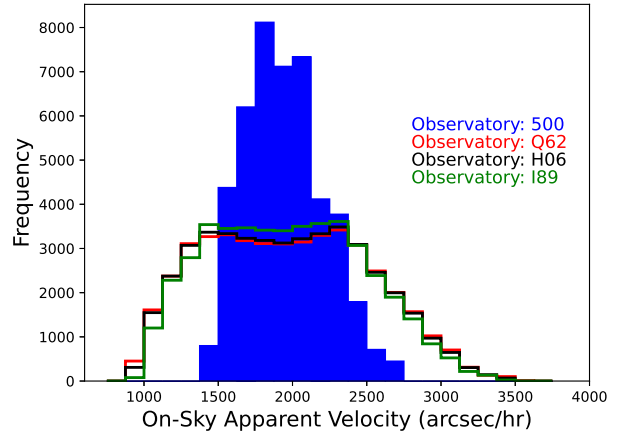}
    \caption{The simulated range of apparent on-sky angular velocities of the potential LTs. The solid blue histogram represents the velocities as seen from the Earth's centre (MPC code: 500). The red step histogram is viewed from the Siding Spring Observatory (Q62), the black step histogram is viewed from Astrocamp (I89), and the green step histogram is viewed from New Mexico Skies (H06). There is a broadening of the range of apparent on-sky angular velocities when observed from the Earth's surface caused by the rotational motion of the observatories around the geocentre.}
    \label{fig:OnSkyRate_AllLocations}
\end{figure}

To obtain a better understanding of the possible on-sky motion of a LT, we used a numerical simulation of hypothetical lunar co-orbital particles to reveal the range of motions expected. The motion of an ideal Trojan asteroid is restricted to the triangular Lagrange points themselves, but real LTs could be almost anywhere along the entirety of the Moon's orbit. To determine optimal search rates for LTs, a set of hypothetical LTs are simulated within a Solar System model which includes the Sun, the 8 planets and the Moon, with initial conditions determined from the JPL DE405 ephemeris \citep{sta98}. One thousand test particles in orbit around the Earth are added, with semimajor axes $a$ within $\pm 30\%$ of the Moon's value, eccentricities $e$ between 0.01 and 0.3, inclinations $i$ of 0 to $45^\circ$, and other angular elements randomized. Most of these are not co-orbital  with the Moon and quickly escape the Earth-Moon system. The 55 which remain near the Moon's orbit after 500 dynamical times (about 40 Earth years) are taken as representative of a temporarily captured co-orbital population. The on-sky motions as observed from different locations on the Earth are used to determine the range of on-sky motions which set the basic parameters of this survey. These can be seen in Figure \ref{fig:OnSkyRate_AllLocations}. From this we found for an observer at the Earth's centre, LTs move at a range of 1400-2700"/hr or 9.33-18.00$^{\circ}/day$, centred on an average of $13.00^\circ/day$.

\subsubsection{The Need for a Dedicated Survey}
Though LTs could be detected by the major NEO surveys, most surveys are looking for asteroids at rates of motion which might cause them to fail to detect LTs, even if these LTs appear in their images. The most likely exception is ATLAS, which does have the capability to detect objects with the necessary range of on-sky speeds. The ATLAS survey may in fact have already effectively done the search for LTs much better than we can here. Yet it is unclear whether their search process --the on-sky search rate of which is not fully detailed in the literature and which may be tuned to the more common and much slower on-sky speeds of typical NEAs-- would detect LTs and with what efficiency. Here we aim to optimize our search process specifically for LTs in an attempt to address the existence of LTs directly.

\section{Methods}

This Lunar Trojan asteroid survey utilizes the iTelescope network \footnote{Accessible online at: \url{https://www.itelescope.net/}} to survey the sky along the Moon's orbit. iTelescope is a not-for-profit organization that allows public access to a series of telescopes via the internet. There are currently 5 observatories around the world on the network, with a total of 20 operational telescopes that range from small refractors up to meter-class reflectors. For our survey, we chose 2 telescopes from each hemisphere, allowing us to take advantage of the best weather across the world throughout the year as well as the position on the sky of the lunar orbital path. All 4 telescopes are relatively small aperture refractors with large FoVs and modest limiting magnitudes: T8 and T9 of the Siding Spring Observatory in New South Wales, Australia (Q62); T16 of Astrocamp in Nerpio, Spain (I89); T20 of New Mexico Skies in Mayhill New Mexico, USA (H06). The properties of these telescopes can be seen in Table \ref{tab:Tele_prop}.
These telescopes were selected by comparing three important properties: FoV, limiting magnitude, and cost per imaging time. From the three properties, these 4 telescopes were determined to be the most cost efficient per square degree while still reaching a significant limiting magnitude in a 30 second exposure.

\begin{table*}
	\centering
	\caption{The properties of the telescopes used in this Lunar Trojan Asteroid Survey. The limiting magnitude is calculated for the 30 second exposure time used here.}
	\label{tab:Tele_prop}
	\begin{tabular}{||c c c c c||} 
		\hline
		Telescope & Aperture ($mm$) & FoV ($arcmin$) & Limiting Magnitude & Pixel Size ($arcsec$) \\
		\hline
		T8 & 106 & 239'x239' & 16.4 & 3.5"x3.5"\\
		T9 & 127 & 189'x189' & 16.4 & 2.75"x2.75"\\
		T16 & 150 & 115'x115' & 16.5 & 1.69"x1.69"\\
		T20 & 106 & 156'x234' & 15.8 & 3.5"x3.5"\\
		\hline
	\end{tabular}
\end{table*}



\subsection{Apparent On-Sky Angular Velocity Range}

The predicted apparent on-sky angular velocities for a LT, as seen from the Earth's centre, is 1400-2700"/hr or 9.33-
18.00$^{\circ}/day$ (solid blue histogram in Figure \ref{fig:OnSkyRate_AllLocations}). However, this range widens when observing from locations on the Earth's surface. The rotation of the Earth creates appreciable motion due to the changing parallax for such nearby asteroids. Therefore, the hypothetical LT simulation was used to determine the on-sky motion for an observer at each of the 3 observatories used in this survey. As expected, the range of the on-sky speeds widens for each location, all in a similar manner.
This effect can be seen in the step histograms in Figure \ref{fig:OnSkyRate_AllLocations}. This provided us with a possible range of 900-3500"/hr or 6.00-23.33$^{\circ}/day$, with a mean of 1970"/hr or $13.13^{\circ}/day$.

\subsection{Survey Technique}

The differentiating factor of this survey is the image cadence. Our cadence is specifically optimized to observe very fast moving objects as they traverse the night sky.

\begin{figure}
	\includegraphics[width=\columnwidth]{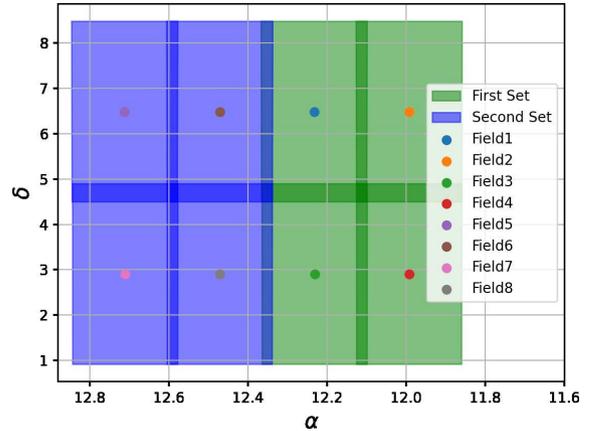}
    \caption{An example of how the Lunar Trojan Asteroid Survey images a 2x4 grid. The first 2x2 set (green) is imaged in chronological order, then repeated for a total of 3 images in each field before moving onto the second set (blue). All adjacent images are overlapped by 10\% of the telescopes FoV to ensure no gaps in the sky coverage of the survey. This example is centred on the Earth-Moon L4 position on January $1^{st}$, 2021 at 00:00 UTC, as seen from the Siding Spring Observatory.}
    \label{fig:2x2Chain}
\end{figure}

This survey collected data in units of 2x2 grids. A 2x2 grid made of 4 fields on the sky is imaged sequentially, then repeated 3 times. This process allows the survey to have repeated images of the same area of the sky quickly, yet leaves enough time between the images for objects such as LTs to move against the background stars. This is important for detecting these fast moving LTs, as too long of a cadence can result in the asteroid moving out of the FoV between images, and therefore becoming undetectable. A full night of surveying consisted of a larger grid centred on a region of the lunar orbital path at the highest altitude on the sky for that night ($>50^\circ$ was preferred). This larger grid was broken down into units of 2x2 grids and once 1 unit was repeated 3 times, the next unit would begin. A graphical representation of the imaging process for a 2x4 grid, split up into 2 units of 2x2 grids is seen in Figure \ref{fig:2x2Chain}. 


A 30 second exposure, captured with an equipped clear ("luminance") filter, is used throughout the survey. This allows the telescopes to reach a reasonable limiting magnitude while still minimizing the amount of trailing that would be produced for objects moving at apparent on-sky angular velocities up to $23.33^{\circ}/day$. A 30 second exposure time on a 2x2 grid of images, with telescope slewing time taken into consideration, results in a typical cadence of 15 minutes between each image of the same area of the sky. Objects moving at 6.00-23.33$^{\circ}/day$ will move $225$-$875$" in these 15 minutes, which translates to $79$-$306$ pixels for our average pixel size. Though optimized for LTs, this cadence makes it possible to expand the searchable on-sky angular velocities range down to lower values, allowing for the detection of other fast moving NEOs. 
For reference, the smallest FoV telescope used in this survey is T16 (115.0'x115.0'). A LT moving at 6.00-23.33$^{\circ}/day$ will traverse the entire FoV in 2.0-7.7 hours.
 
\subsection{Apparent Magnitude and Asteroid Size Estimates}

The apparent magnitude of these theoretical LTs is calculated from a routinely used set of formulas that can convert an apparent magnitude to a diameter using the well-known HG formalism outlined by \cite{Bowell_1989} and the equation for diameter outlined by \cite{Harris_1997}.

\begin{figure}
	\includegraphics[width=\columnwidth]{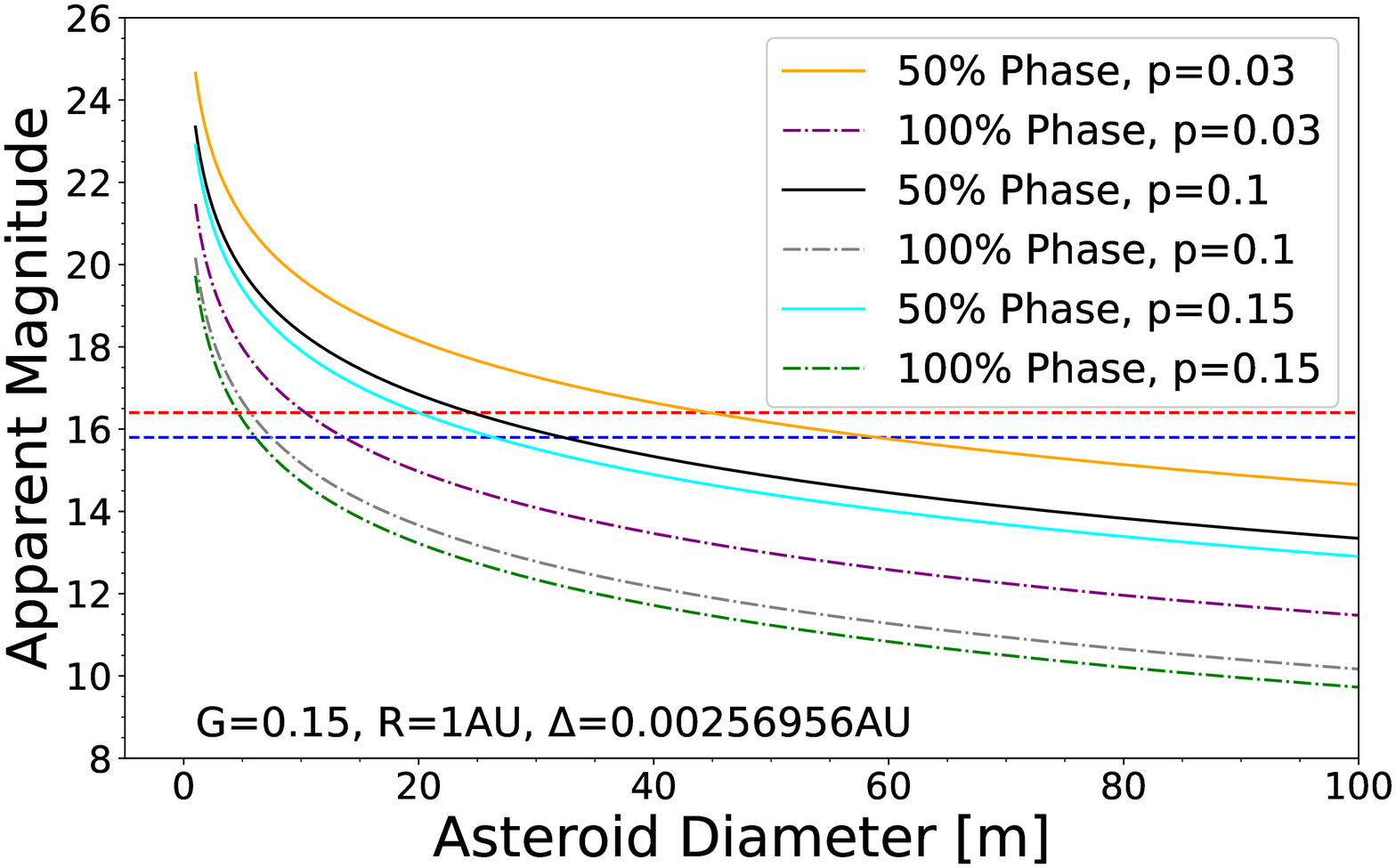}
    \caption{The apparent magnitude of the theoretical LTs, calculated as a function of diameter [$m$] at $50\%$ and $100\%$ phases for albedos of $p=0.03$, $p=0.1$, and $p=0.15$. Assumed values are: $G=0.15$, $R=1\;[AU]$, and  $\Delta=0.00256956\;[AU]$. The limiting magnitude of the 4 telescopes used in this survey are also plotted to visualize our capabilities. The red dashed line is the limiting magnitude of T8, T9, and T16 (16.4) while the blue dashed line is the limiting magnitude of T20 (15.8).}
    \label{fig:appmag_alb}
\end{figure}

We assume the values: slope parameter $G=0.15$, distance from the Sun $R=1~[AU]$ and distance from the Earth $\Delta= 0.00256956~[AU]$ (the average LD). With these values, the apparent magnitude of the theoretical LTs is plotted in Figure \ref{fig:appmag_alb} as a function of asteroid diameter, from 1-100$\;[m]$, using a phase angle of $0$ and $\pi/2$ ($100\%$ and $50\%$ illumination phases respectively), with varying albedo values of typical of NEOs: 0.03, 0.1 and 0.15 \citep{Hartmann_2005}. Overlaid on this plot is the limiting magnitude of the telescopes for a 30 second exposure, illustrating our survey's capability of detecting LTs down to a size of approximately $5\;[m]$ in diameter when fully illuminated with an albedo of 0.15. At lower albedos and illumination phases, the size limit increases, reaching roughly $60\;[m]$ at $50\%$ phase with an albedo of 0.03.

\subsection{Data collection and processing}

iTelescope images were automatically bias, dark and flat field corrected. These calibrated images were downloaded from their server and plate solved using the PinPoint Astrometric Engine \footnote{Available online at: \url{http://pinpoint.dc3.com}} from  DC-3 Dreams. Once completed, the images are ready for processing in the search for moving sources.

The search for moving sources is done with a Python code used in previous work for main-belt asteroid detection in archival telescope images (\cite{Wie07}, \cite{GilWie09}, \cite{GilWie10}, and \cite{AugWie13}). This code takes a triplet of images of the same area of the sky and uses the software package Source Extractor \citep{SourceExt} to locate the sources in each image. Stationary sources are removed, and the remaining list is examined for moving sources consistent with the range of on-sky motions being searched for. Candidate moving sources are presented to a human operator for verification.

Here we require that the source is moving in a nearly-straight path at a nearly-constant rate of motion within any image triplet.
Specifically, the motion from the first to the second image was taken to define a straight line; and then the on-sky motion of the source between the second and third images had to land at the nominal location within a tolerance corresponding to 10\% of the expected on-sky speed.

Our set minimum and maximum apparent angular velocities were 300-3000"/hr, or 2-20$^{\circ}/day$. This range of apparent angular velocities extends below the $6^\circ/day$ expected minimum on-sky rate of motion of a LT. This broadens the surveys capabilities beyond LTs to more common NEAs. \cite{Tricarico17} states that 99\% of NEAs have apparent on-sky angular velocities below $10^\circ/day$ and 90\% below $3^\circ/day$. As a result, our survey is capable of detecting some fraction of the fastest NEAs. We also note that according to the Jet Propulsion Laboratory's (JPL) HORIZONS \footnote{Accessible at: \url{https://ssd.jpl.nasa.gov/horizons.cgi}}, the minimoons 2006 RH$_{120}$ and 2020 CD$_3$ had ranges of on-sky apparent angular velocities of 0.3-61.5$^\circ/day$ and 0.1-27.9$^\circ/day$ respectively, when captured by Earth. Although the minimum and maximum ranges of motion for these minimoons are beyond the capabilities of this survey, our survey does heavily overlap their on-sky rates of motion.
We note that our software's maximum detectable velocity threshold (3000"/hr) was set below the maximum simulated on-sky velocity (3500"/hr, Figure \ref{fig:OnSkyRate_AllLocations}). This was needed to limit the survey software false detection rate, though it does exclude a small fraction of the possible on-sky speeds.

\subsubsection{Detection Efficiency Test}

To properly test the efficiency of the moving source detection software, artificial asteroids with randomized rates and directions of motion and magnitudes were implanted into real images taken by each of the telescopes. These images were then searched for the seeded sources using the detection software. To accurately model the motion of LTs on and across the images, proper trailing and deviation from straight line-of-motion (curvature) was accounted for.

The trailing was accounted for by creating trailed artificial sources for implantation. This was accomplished by co-adding the Gaussian source corresponding to a 1 second exposure in 30 consecutive locations according to the rate of motion for the random artificial asteroid.

The curvature of the LTs was established from the numerical simulations used previously in section \ref{sec:Simulation} to determine the expected on-sky motion. The deviation from a straight line was typically $\pm10$" in Declination and $\pm20$" in Right Ascension during the image sequence. These correspond to on-sky speed variations of a few percent, within the 10\% tolerances set in the software. Curvatures at these levels were introduced randomly into our artificially implanted images to properly account for its effect on our detection pipeline.

Each telescope was tested by implanting 1000 artificial asteroids moving at random on-sky rates of motion between 300-3000"/hr with random magnitudes between 8.0-18.0. The seeded images were then run through the moving source detection software.

The detection efficiency as a function of apparent magnitude is shown in detail in Appendix \ref{ArtificialPlots}.
For T8, the limiting magnitude was determined to be $\approx16.5$. For T9, it was 16. For T20, it was 15. There is no effect of apparent on-sky angular velocity on detection efficiency in these three telescopes.

For T16, the limiting magnitude was determined to be $\approx15.5$. In this case, there is an observed decrease in detection efficiency with apparent on-sky angular velocity. This is likely caused by the small pixel size (1.69") causing additional trailing losses, and the slow telescope slewing, causing a longer typical cadence. This effect is most important for LTs with on-sky angular velocities $>2000$"/hr.

The detailed results for each telescope can be seen in the histograms of Appendix \ref{ArtificialPlots}.

\section{Results and Discussion}



\subsection{Survey Results}

\begin{figure*}
 \begin{center}
\includegraphics[width=\textwidth]{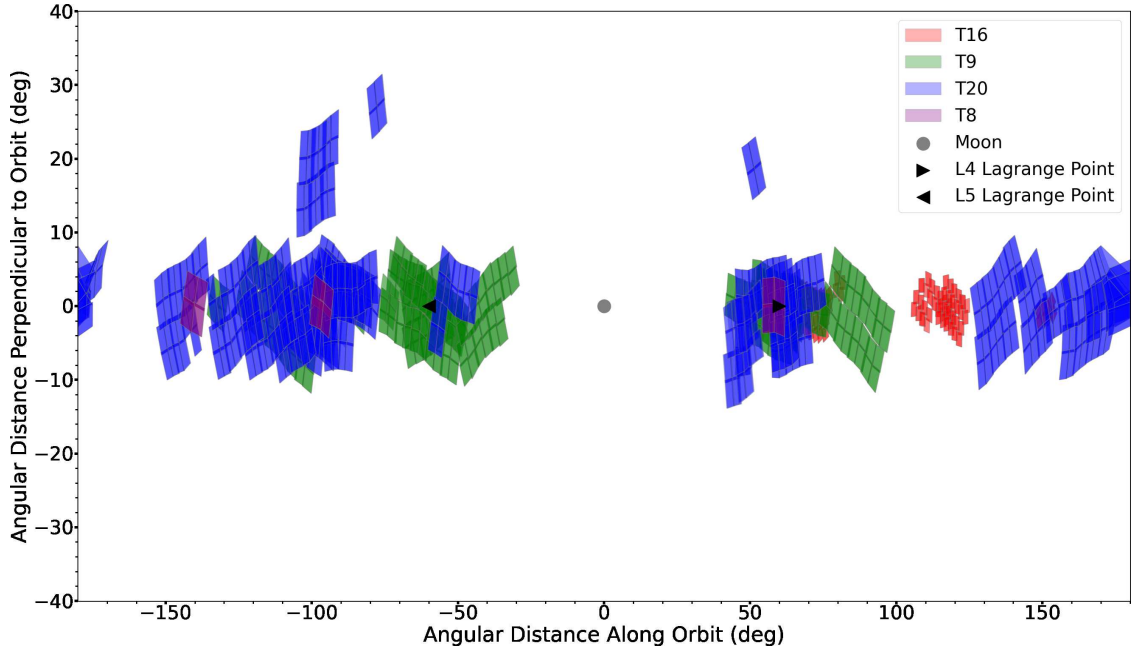}
\caption{The total coverage of the Lunar Trojan Asteroid Survey, relative to the on-sky lunar orbital path. In this depiction, the Moon remains at the point (0,0) for all times. The L4 and L5 Lagrange points are illustrated by triangles, clearly displaying the surveys coverage of these particular regions. The images are colour coded to represent the telescope the data was obtained by.}
\label{fig:LunarCov}
\end{center}
\end{figure*}

The survey was conducted for 12 months, resulting in 882 fields of 3 images that were searched for moving sources. The coverage of the survey, with respect to the lunar orbit, can be seen in Figure \ref{fig:LunarCov}. The survey covered areas of the lunar orbit extending from an angular distance of $\pm30^\circ$ to $\pm180^\circ$ from the Moon along its orbit.

The 882 fields the survey captured totalled $8340\;deg^2$. If this total survey area is compared to the area of the the entire celestial sphere, the survey amounts to nearly 20\% coverage. In practice, the survey focused along the lunar orbit, repeating some regions more than once.

The region of interest is that $\pm10^\circ$ latitude from the on-sky lunar path ($7200\;deg^2$ ). Neglecting the region with  $\pm30^\circ$ of the Moon, which cannot be effectively searched due to lunar interference, leaves $6000\;deg^2$. Our survey area concentrated on this region and imaged 76\% of it ($4570\;deg^2$) at least once. 

No LTs were discovered during this survey.  Given the rapid rate that LTs are expected to move on the sky, and that our survey images covered an area roughly 20\% of the total celestial sphere, we estimate $\lesssim 5$ asteroids that are co-orbital with the Moon at the present time, down to our limiting magnitude, between 15.0-16.0. A more precise limit on the number of LTs would require more careful modelling of the expected LT directions and speeds of motion relative to our actual coverage and image timing; nevertheless our estimate is likely accurate to a factor of 2.

\begin{figure}
    \includegraphics[width=\columnwidth]{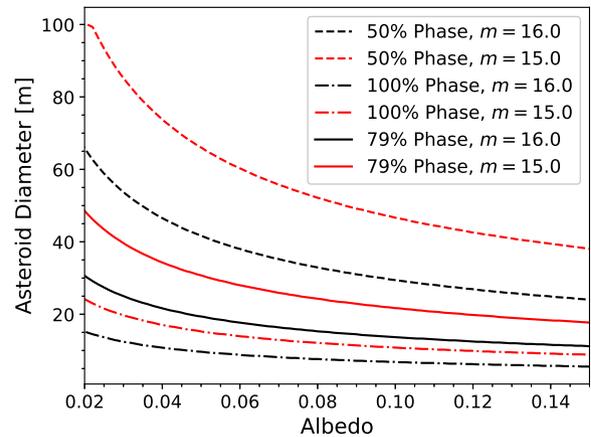}
    \caption{Asteroid diameter as a function of albedo for a theoretical LT at an apparent magnitude of 15.0 (red)} and 16.0 (black), each when seen at $50\%$ (dashed line), $79\%$ (solid line) and $100\%$ (dashdot line) illumination phases. Assumed values are: $G=0.15$, $R=1\;[AU]$, and  $\Delta=0.00256956\;[AU]$.
    \label{fig:LunarDiam_alb}
\end{figure}

A 15-$16^{th}$ magnitude asteroid (which is equivalent to an asteroid with absolute magnitude $H \approx 26.4$-$27.4$ under our assumed survey conditions) corresponds to a range of sizes given the possible albedos and phases of observation. As seen in Figure \ref{fig:appmag_alb}, our minimum observable diameter ranges from roughly 5-60$\;[m]$.  Figure \ref{fig:LunarDiam_alb} compares asteroid diameter and albedo for $50\%$ and $100\%$ illumination phases, as well as the survey's expected average phase of 79\%, for a 15.0 and 16.0 magnitude object. For a typical NEA albedo of 0.08 \citep{RyanWoodward10}, the corresponding size is $15\;[m]$ in diameter for a 16.0 magnitude object, and $24\;[m]$ for a 15.0 magnitude object. Our survey cannot effectively set a limit on the number of LTs on the Moon's orbit at smaller sizes.

For a conclusive estimate, we use our lower end of the limiting magnitude. We then estimate $\lesssim 5$ asteroids that are co-orbital with the Moon at the present time, down to our limiting magnitude of 15.0, and therefore with $H<26.4$. For an albedo of 0.08, this corresponds to objects with diameters $>24\;[m]$.

\subsection{NEA Detection - \textit{ALA2xH}}

\begin{figure*}
    \centering
    \includegraphics[width=5.905cm]{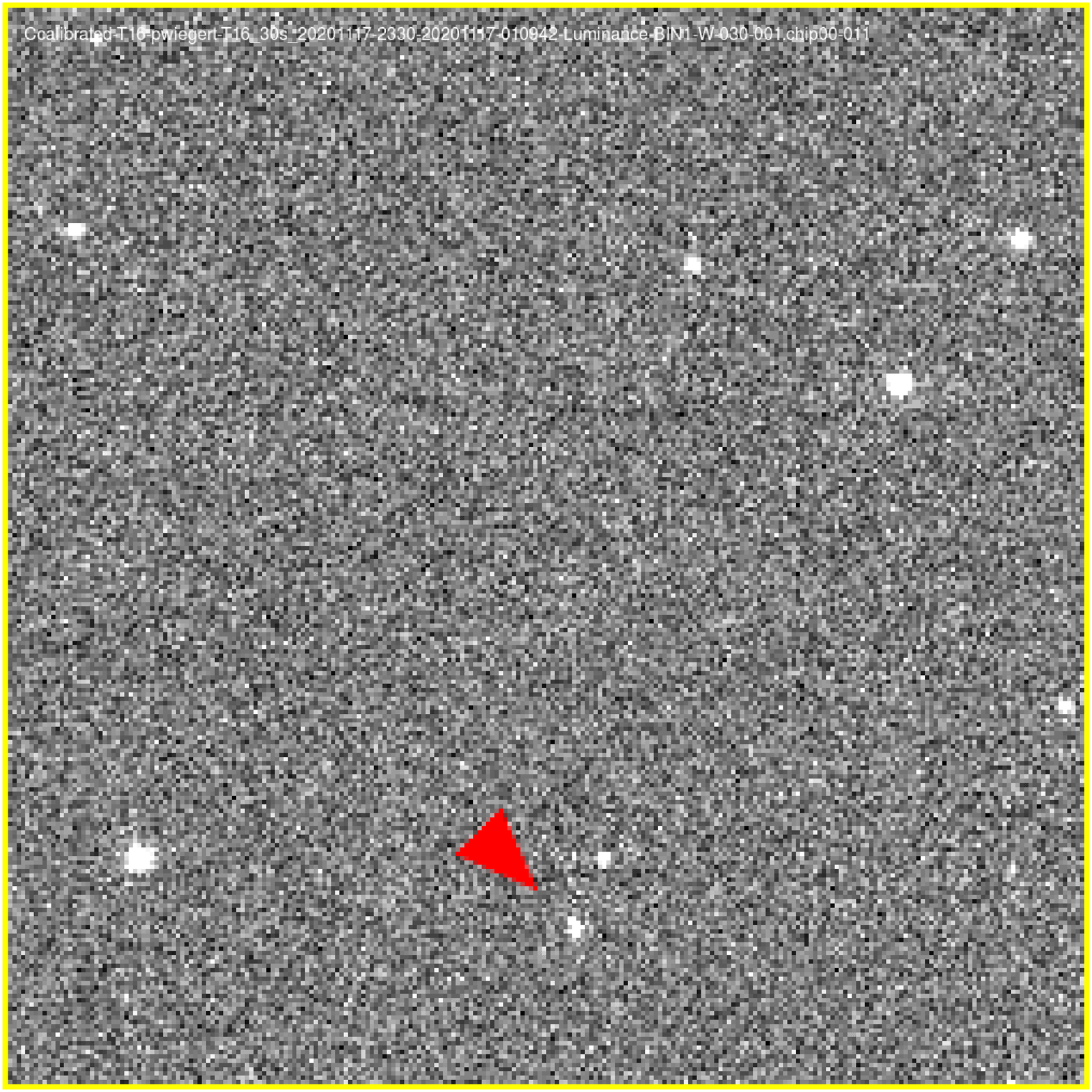}
    \includegraphics[width=5.905cm]{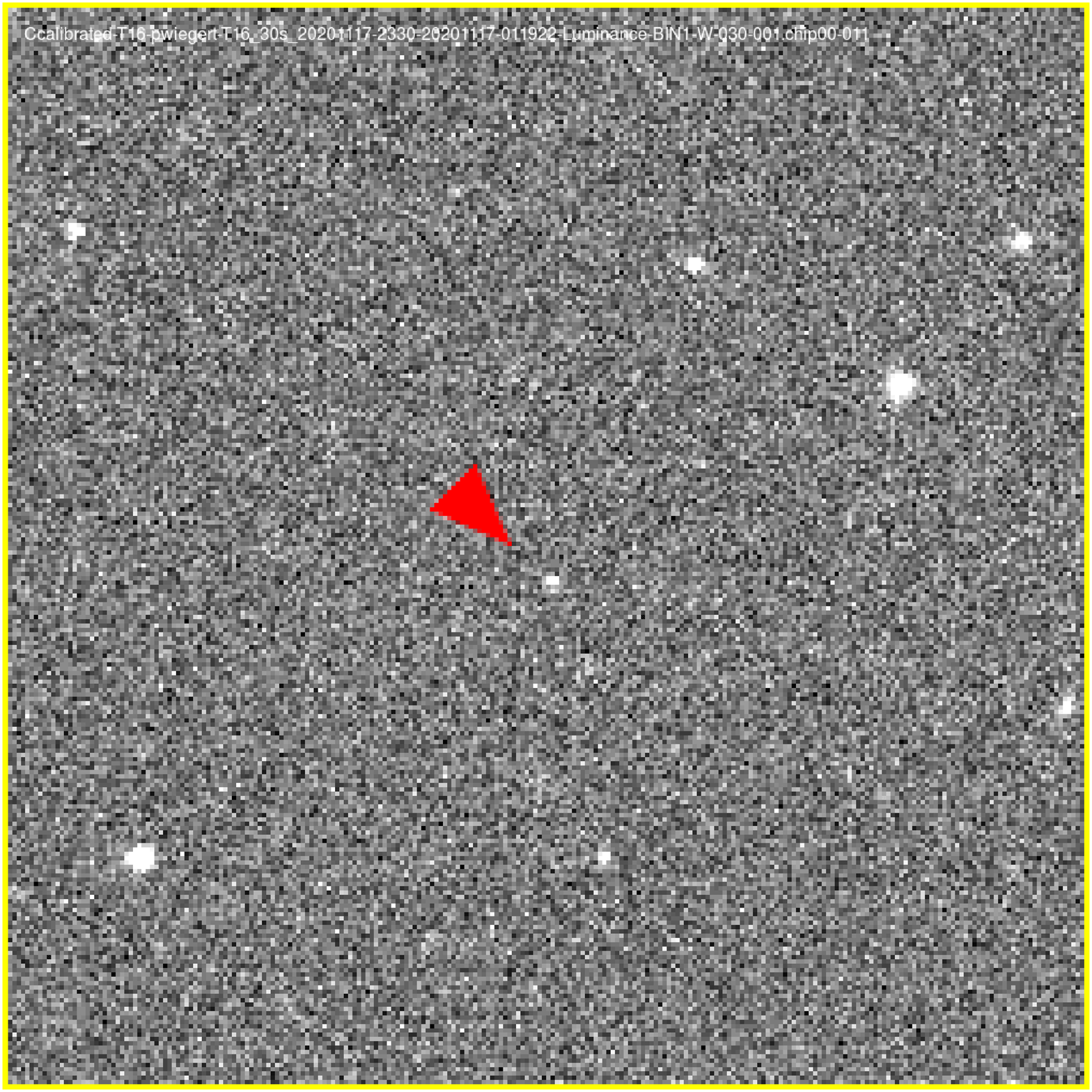}
    \includegraphics[width=5.905cm]{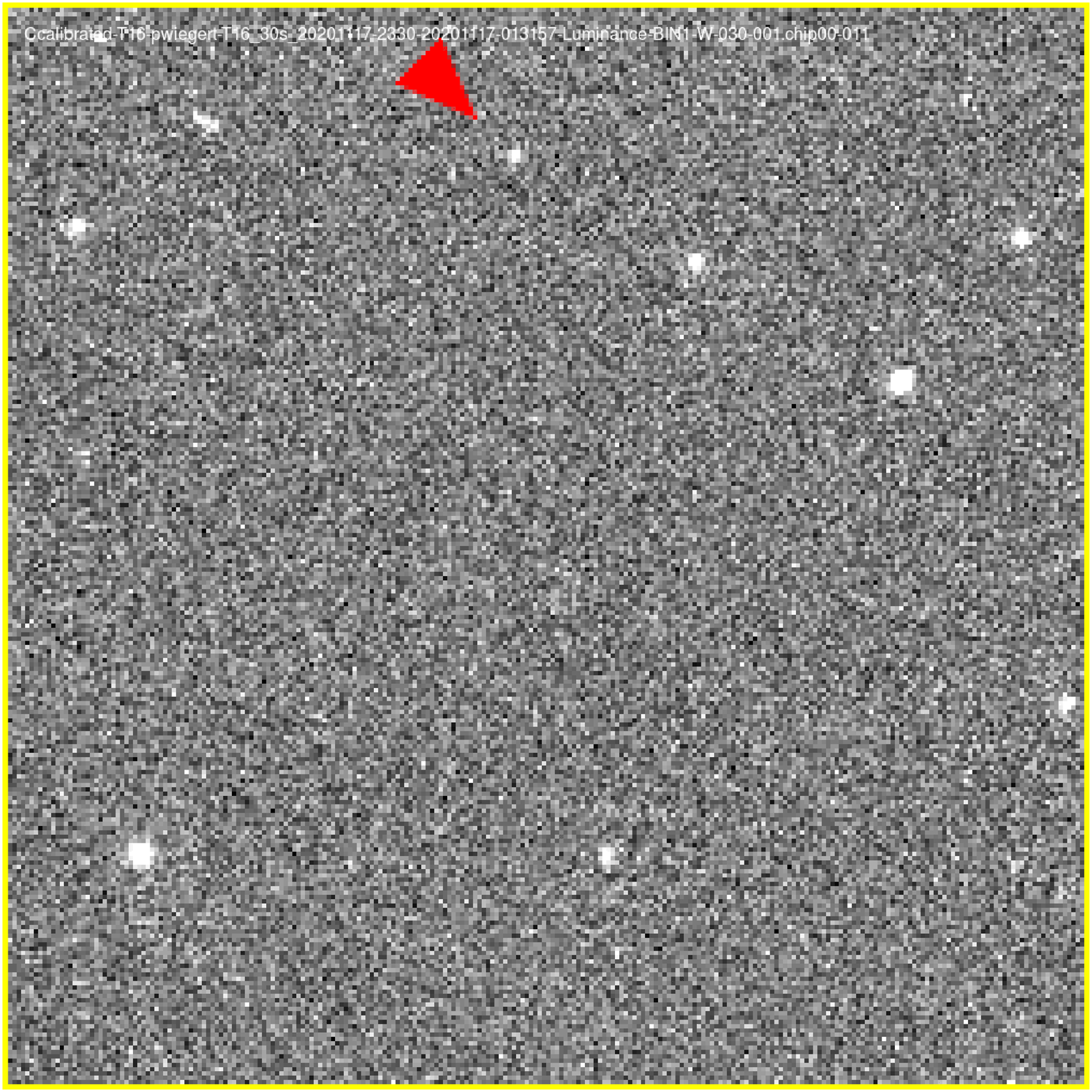}
\caption{The discovery images of ALA2xH, a 15.0 magnitude asteroid moving at 810"/hr, observed by T16 in Nerpio, Spain at the Astrocamp Observatory. The image sequence starts at 23:09:42 UTC on November $17^{th}$, 2020 with time intervals of 9 minutes 22 seconds and 12 minutes 35 seconds.}
\label{fig:ALA2xH}
\end{figure*}

Whilst the survey did not see any LTs, on the night of November $17^{th}$, 2020 a detection was made of a 15.0 magnitude object, moving at 810"/hr or $5.4^\circ/day$ over a 0.37 hour arc. This observation is shown in Figure \ref{fig:ALA2xH}. Known objects in the same region of the sky were checked through the MPC's MPChecker\footnote{Available online at: \url{https://minorplanetcenter.net/cgi-bin/checkmp.cgi}} and NEOChecker\footnote{Available online at: \url{https://minorplanetcenter.net/cgi-bin/checkneo.cgi}}, with no positive match. The observation was submitted to the MPC, and follow-up images of the object were attempted, but unfortunately these were unsuccessful due to poor weather conditions and telescope availability. The MPC gave the object a temporary designation \textit{ALA2xH}.

When reported to the MPC, JPL's Scout \footnote{Available online at: \url{https://cneos.jpl.nasa.gov/scout/intro.html}} provides a trajectory analysis and hazard assessment of the observed object, including a likelihood score of it being a NEO, PHA, geocentric, cometary, and having an interior-to-Earth orbit (IEO). Scout gave the observation of \textit{ALA2xH} a 100\% likelihood score of the object being a NEO, 5\% of being a PHA, 0\% of being geocentric, 40\% of being cometary, and 0\% of having an IEO. Scout also gave \textit{ALA2xH} a modest risk of impacting Earth (a 2 on a scale of 0-4) It is estimated to be $H=24.3$ (50-100$\;[m]$ in diameter) with a most probable closest approach of 0.05 LD, or $19,220\;[km]$ ($\approx 3$ Earth-radii). Though no follow-up observations were obtained, by us or by others, this detection is a substantial proof of concept for this survey.

\section{Conclusions}

This Lunar Trojan asteroid survey succeeded in surveying much of the Moon's orbit during the 12 months of its operation. The survey was designed to detect Lunar Trojans (LTs) and other bodies moving with apparent on-sky angular velocities of 2-20$^\circ/day$ down to the $15^{th}$ magnitude. No LTs were discovered. 

Our survey did however detect one fast-moving NEA. This $H = 24.3$ asteroid, with the temporary designation ALA2xH, had a close approach with Earth, having a most probable miss distance of 0.05 LD ($\approx 3$ Earth-radii). The asteroid was moving below the minimum expected motion of a LT when observed but did elude the major surveys. 

From our analysis, we can set an upper limit of $\lesssim 5$ LTs that currently exist with $H<26.4$. For our average observed phase of 79\%, this corresponds to asteroids $>40\;[m]$ for an albedo of 0.03, $>24\;[m]$ for an albedo of 0.08, and $>18\;[m]$ for an albedo of 0.15. 

The value of asteroids, particularly nearby ones, will only grow as humanity becomes increasingly active in near-Earth space. Though LTs may not be a particularly abundant population, they may still be present in small numbers, only awaiting detection to unlock their potential. Larger and more sophisticated surveys for them are encouraged.





\section*{Acknowledgements}

Funding for this work was provided in part by the Natural Sciences and Engineering Research Council of Canada Discovery Grants program (Grant no. RGPIN-2018-05659). Thank you to the team at iTelescope.Net for making the telescopes available to the public, allowing this project to be conducted.

\section*{Data Availability}

The data analysed in this work was collected as a part of this project and can be made available upon reasonable request. For more information, please contact the first author: cgregg2@uwo.ca.
 



\bibliographystyle{mnras}
\bibliography{main.bib} 



\appendix

\section{Detection Efficiency Plots}\label{ArtificialPlots}

Histograms displaying the artificial asteroids results from each of the telescopes used in this survey. Each telescope has 2 plots associated with it, one presenting the magnitude of the detections compared to the implanted magnitudes, and one for the detected and implanted apparent on-sky angular velocities.

\begin{figure*}
	\includegraphics[width=\columnwidth]{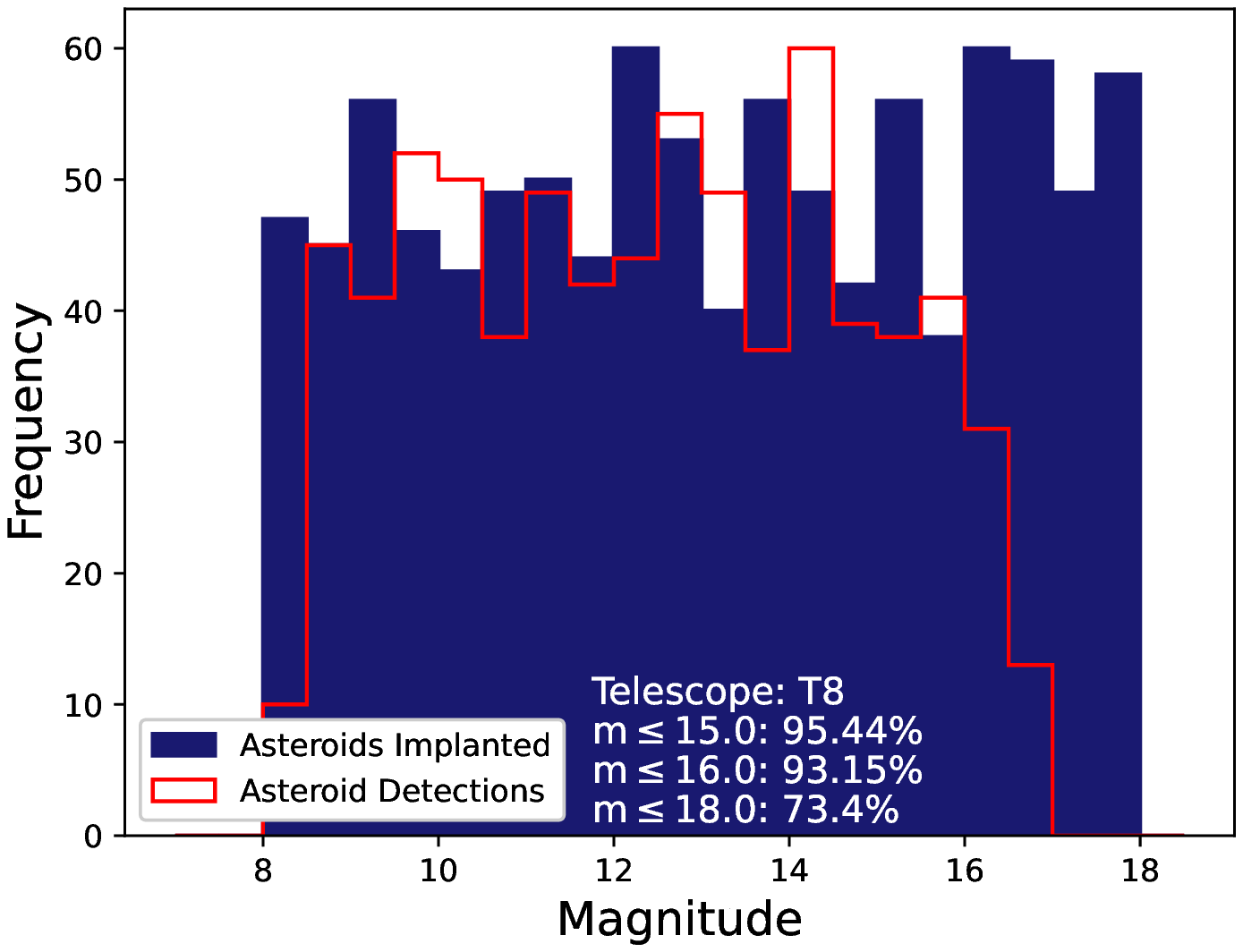}
	\includegraphics[width=\columnwidth]{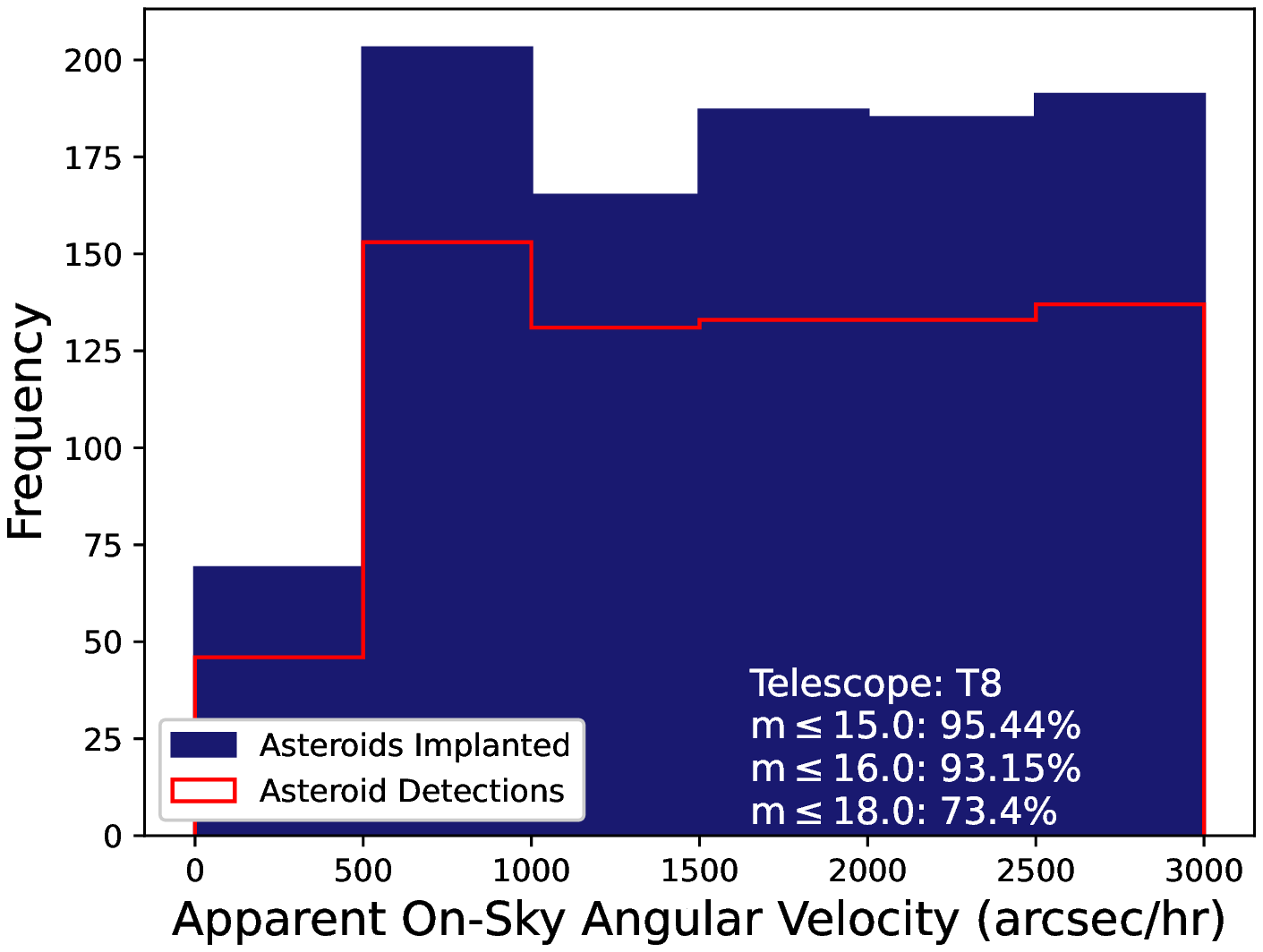}
    \caption{The left panel displays the magnitudes of the 1000 implanted artificial asteroids on a T8 image displayed with the solid blue histogram, overlayed with the red step histogram of the detected magnitudes. The right panel displays the apparent on-sky angular velocities of the 1000 implanted artificial asteroids on a T8 image displayed with the solid blue histogram, overlayed with the red step histogram of the detected velocities. The result gives a 95\% detection rate for asteroids with apparent magnitudes $\leq15.0$ and 83\% with apparent magnitude $\leq16.0$. The limiting magnitude is $\approx16.5$. There is no effect of apparent on-sky angular velocity on detection efficiency.}
    \label{fig:Artif_T8}
\end{figure*}

\begin{figure*}
	\includegraphics[width=\columnwidth]{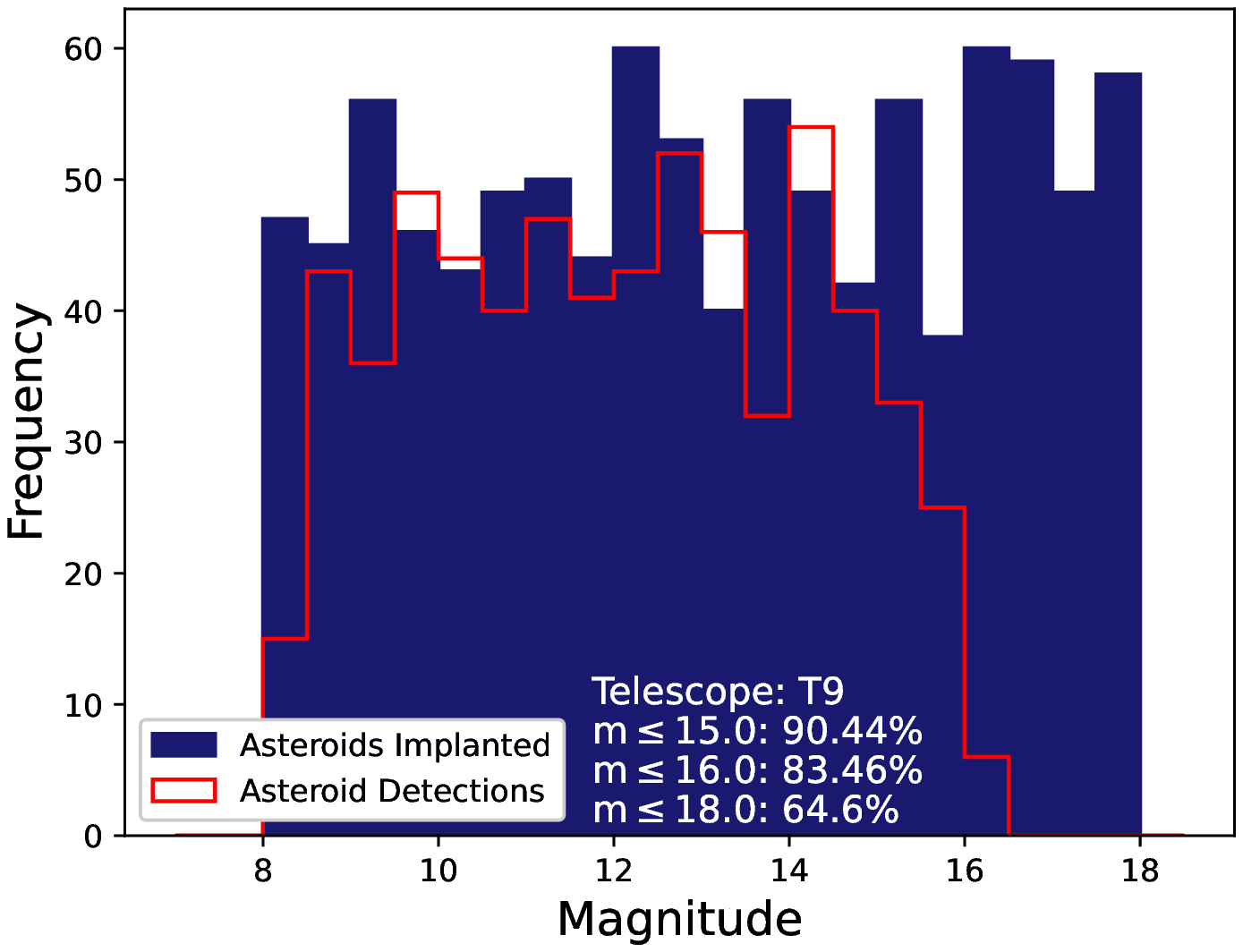}
	\includegraphics[width=\columnwidth]{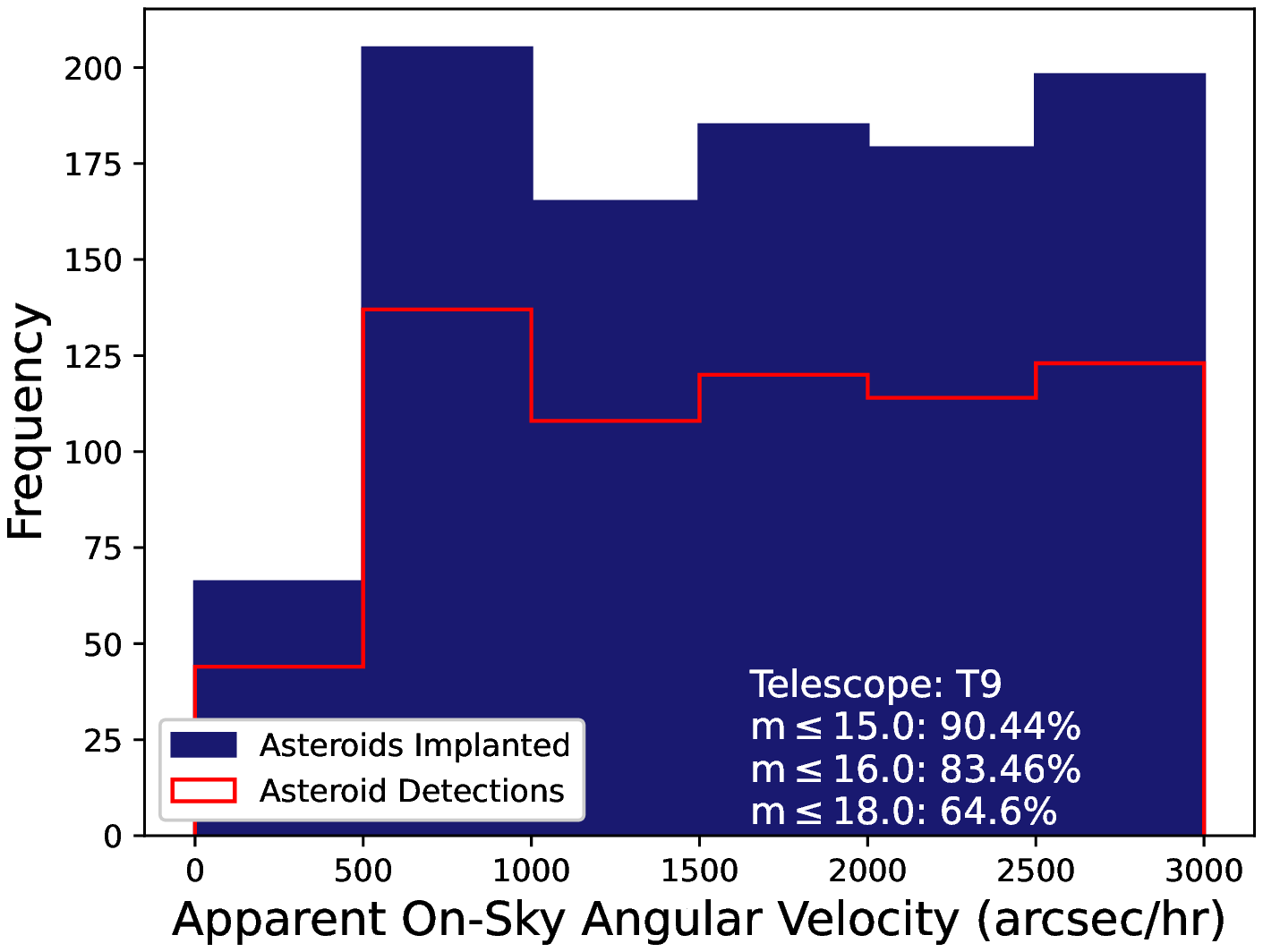}
    \caption{The left panel displays the magnitudes of the 1000 implanted artificial asteroids on a T9 image displayed with the solid blue histogram, overlayed with the red step histogram of the detected magnitudes. The right panel displays the apparent on-sky angular velocities of the 1000 implanted artificial asteroids on a T9 image displayed with the solid blue histogram, overlayed with the red step histogram of the detected velocities. The result gives a 90\% detection rate for asteroids with apparent magnitudes $\leq15.0$ and 83\% with apparent magnitude $\leq16.0$. The limiting magnitude is $\approx16.0$. There is no effect of apparent on-sky angular velocity on detection efficiency.}
    \label{fig:Artif_T9}
\end{figure*}

\begin{figure*}
	\includegraphics[width=\columnwidth]{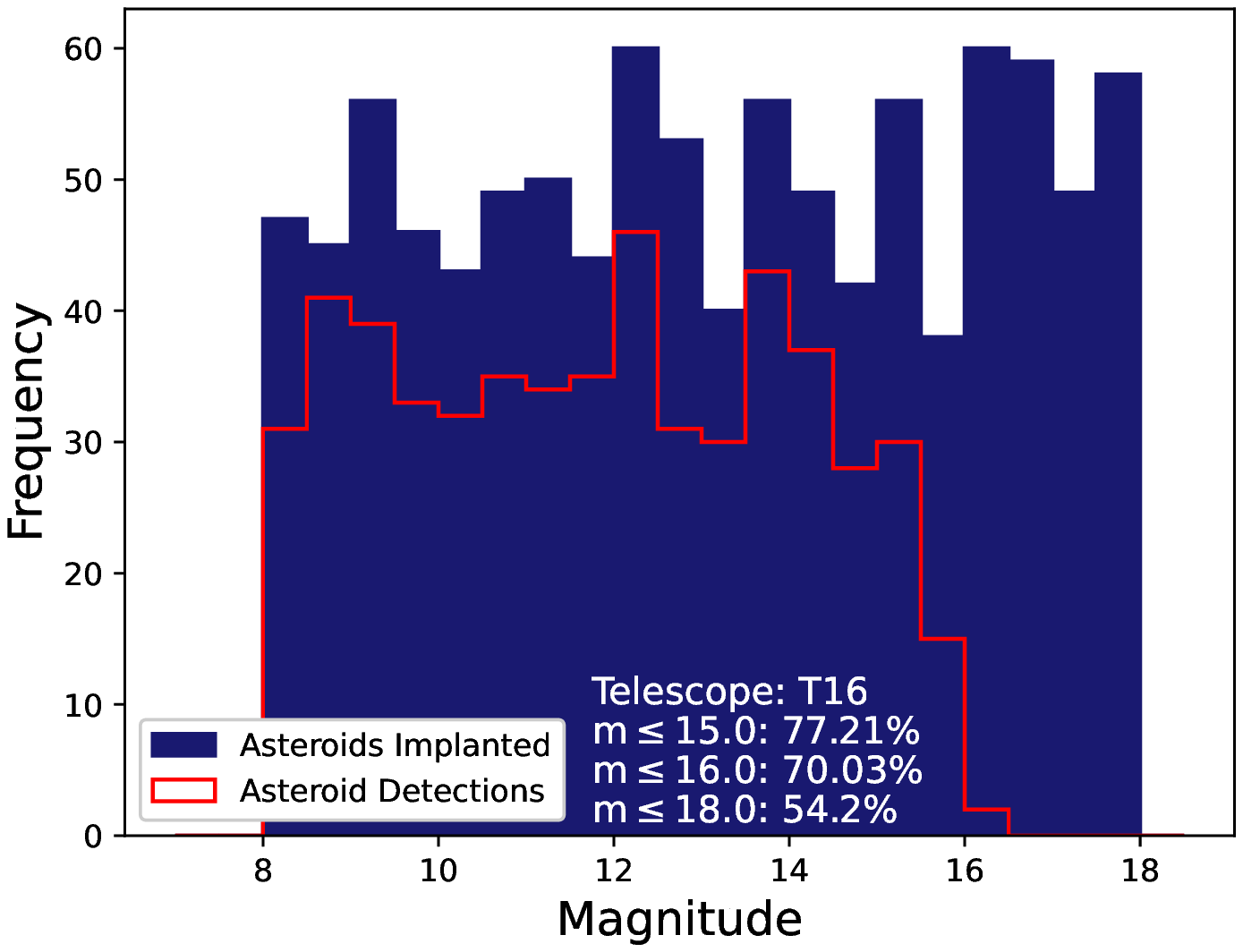}
	\includegraphics[width=\columnwidth]{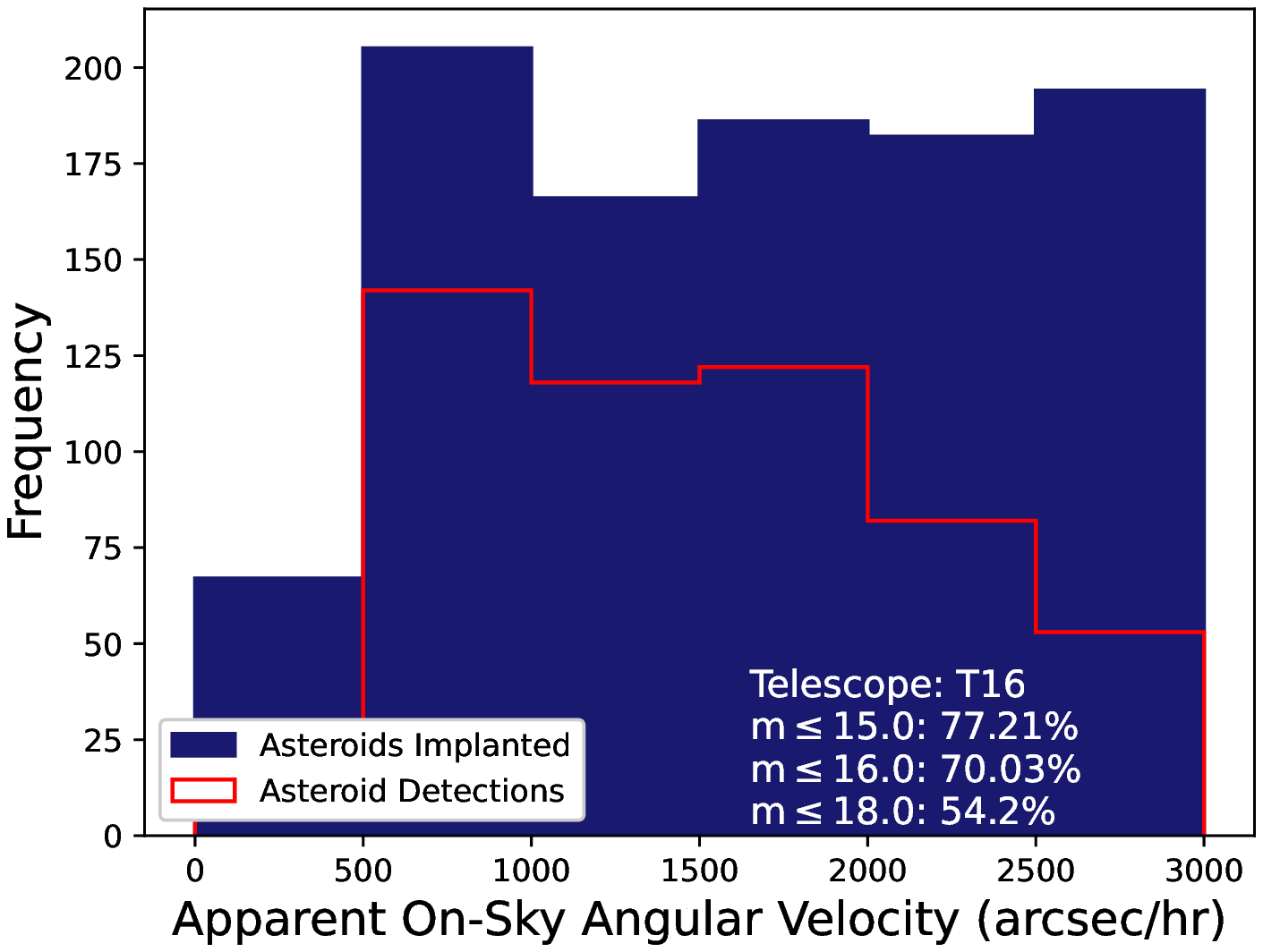}
    \caption{The left panel displays the magnitudes of the 1000 implanted artificial asteroids on a T16 image displayed with the solid blue histogram, overlayed with the red step histogram of the detected magnitudes. The right panel displays the apparent on-sky angular velocities of the 1000 implanted artificial asteroids on a T16 image displayed with the solid blue histogram, overlayed with the red step histogram of the detected velocities. The result gives a 77\% detection rate for asteroids with apparent magnitudes $\leq15.0$ and 70\% with apparent magnitude $\leq16.0$. The limiting magnitude is $\approx15.5$. There is an effect of apparent on-sky angular velocity on detection efficiency seen for T16. This is expected to be caused by the small pixel size (1.69") and the slow telescope slewing, causing a longer typical cadence.}
    \label{fig:Artif_T16}
\end{figure*}

\begin{figure*}
	\includegraphics[width=\columnwidth]{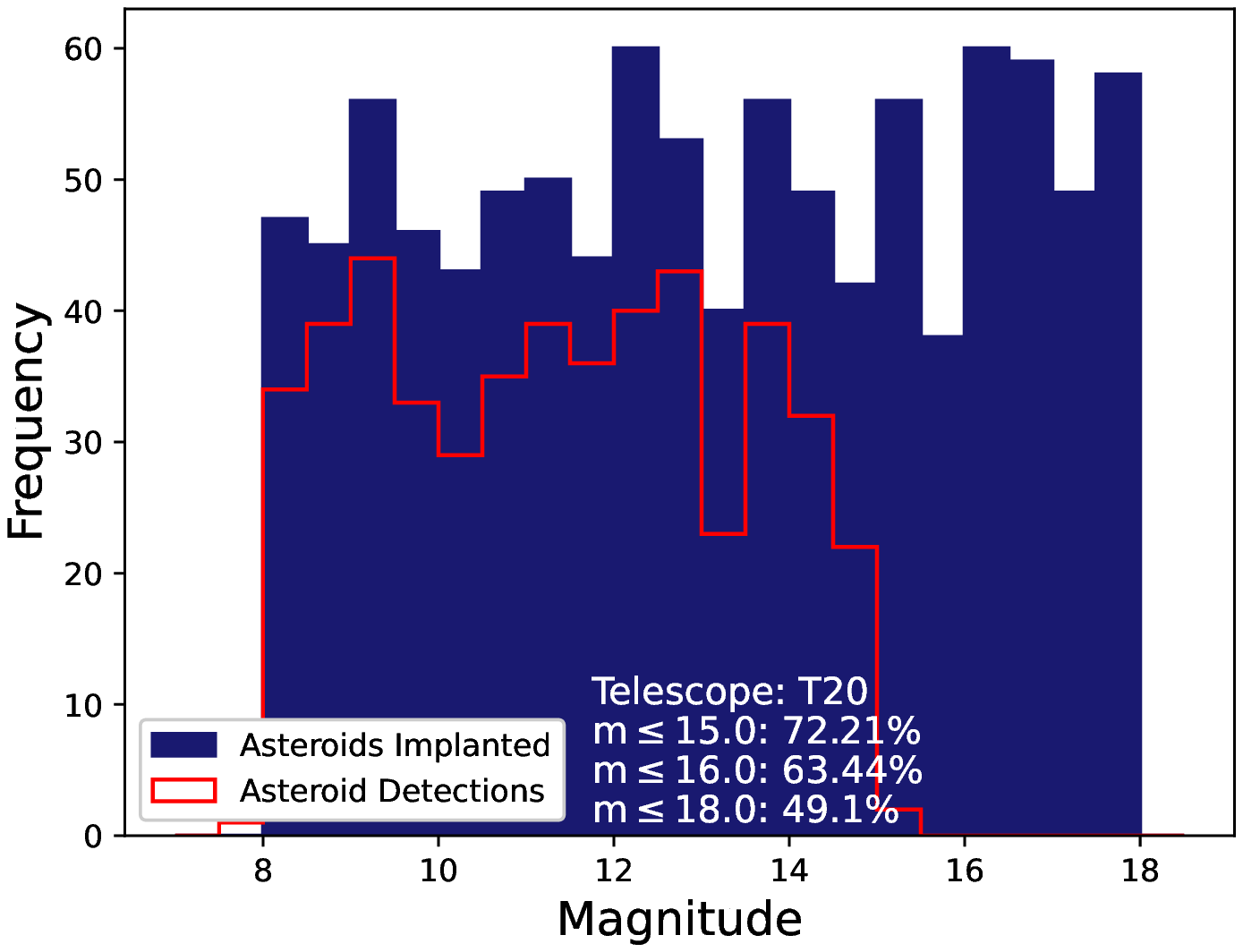}
	\includegraphics[width=\columnwidth]{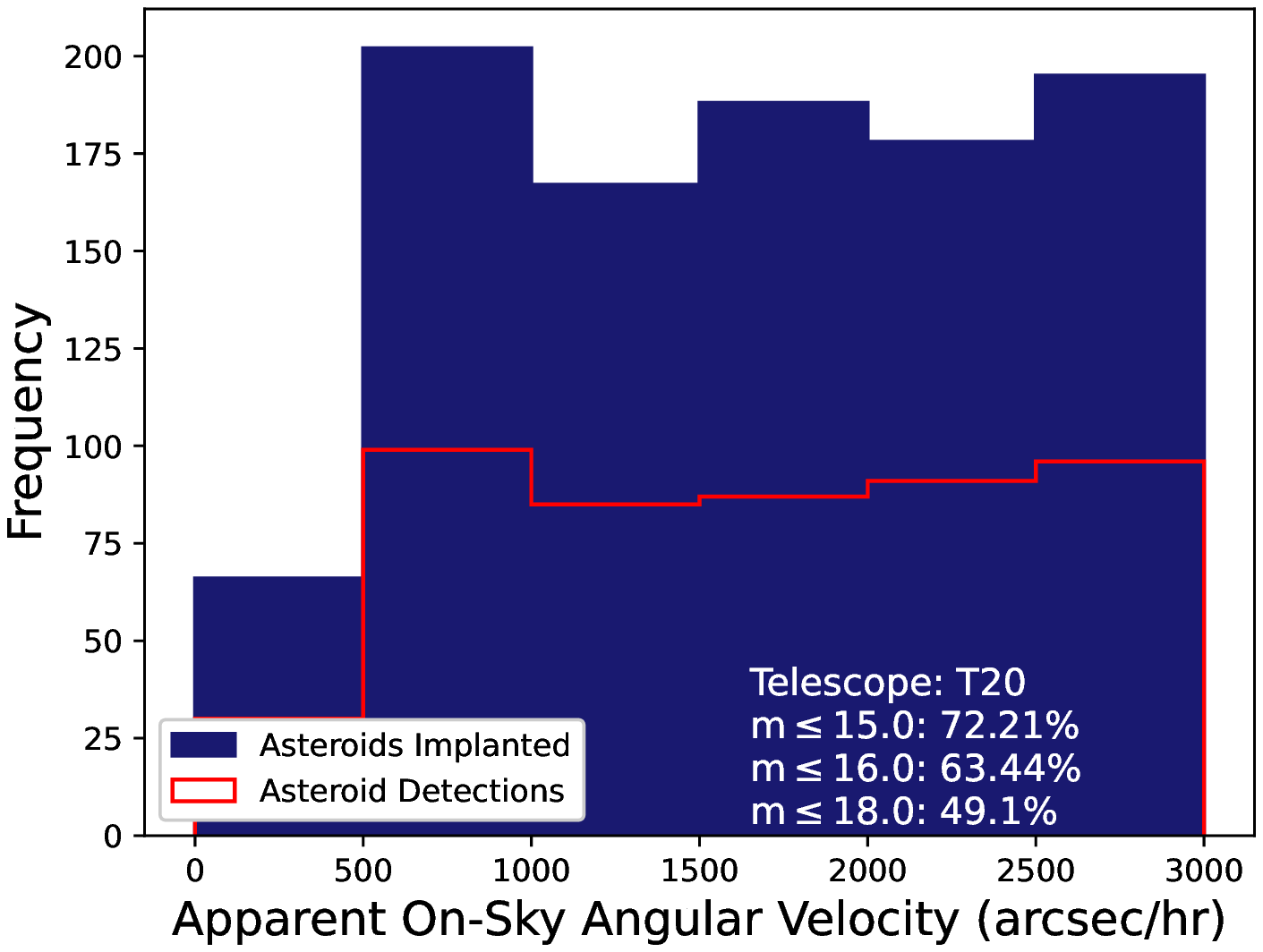}
    \caption{The left panel displays the magnitudes of the 1000 implanted artificial asteroids on a T20 image displayed with the solid blue histogram, overlayed with the red step histogram of the detected magnitudes. The right panel displays the apparent on-sky angular velocities of the 1000 implanted artificial asteroids on a T20 image displayed with the solid blue histogram, overlayed with the red step histogram of the detected velocities. The result gives a 72\% detection rate for asteroids with apparent magnitudes $\leq15.0$ and 63\% with apparent magnitude $\leq16.0$. The limiting magnitude is $\approx15.0$. There is no effect of apparent on-sky angular velocity on detection efficiency.}
    \label{fig:Artif_T20}
\end{figure*}






\bsp	
\label{lastpage}
\end{document}